\documentclass[aps,prl,twocolumn,showpacs,superscriptaddress]{revtex4}
\usepackage{graphicx}

\begin{document}

\title{Nematic Quantum Critical Fluctuations in BaFe$_{2-x}$Ni$_x$As$_2$}

\author{Zhaoyu Liu}
\affiliation{Beijing National Laboratory for Condensed Matter Physics, Institute of Physics, Chinese Academy of Sciences, Beijing 100190, China}
\author{Yanhong Gu}
\affiliation{Beijing National Laboratory for Condensed Matter Physics, Institute of Physics, Chinese Academy of Sciences, Beijing 100190, China}
\author{Wei Zhang}
\affiliation{Beijing National Laboratory for Condensed Matter Physics, Institute of Physics, Chinese Academy of Sciences, Beijing 100190, China}
\author{Dongliang Gong}
\affiliation{Beijing National Laboratory for Condensed Matter Physics, Institute of Physics, Chinese Academy of Sciences, Beijing 100190, China}
\author{Wenliang Zhang}
\affiliation{Beijing National Laboratory for Condensed Matter Physics, Institute of Physics, Chinese Academy of Sciences, Beijing 100190, China}
\author{Tao Xie}
\affiliation{Beijing National Laboratory for Condensed Matter Physics, Institute of Physics, Chinese Academy of Sciences, Beijing 100190, China}
\author{Xingye Lu}
\affiliation{Beijing National Laboratory for Condensed Matter Physics, Institute of Physics, Chinese Academy of Sciences, Beijing 100190, China}
\author{Xiaoyan Ma}
\affiliation{Beijing National Laboratory for Condensed Matter Physics, Institute of Physics, Chinese Academy of Sciences, Beijing 100190, China}
\author{Xiaotian Zhang}
\affiliation{Beijing National Laboratory for Condensed Matter Physics, Institute of Physics, Chinese Academy of Sciences, Beijing 100190, China}
\author{Rui Zhang}
\affiliation{Department of Physics and Astronomy, Rice University, Houston, Texas 77005-1827, USA}
\author{Jun Zhu}
\affiliation{Beijing National Laboratory for Condensed Matter Physics, Institute of Physics, Chinese Academy of Sciences, Beijing 100190, China}
\author{Cong Ren}
\affiliation{Beijing National Laboratory for Condensed Matter Physics, Institute of Physics, Chinese Academy of Sciences, Beijing 100190, China}
\author{Lei Shan}
\affiliation{Beijing National Laboratory for Condensed Matter Physics, Institute of Physics, Chinese Academy of Sciences, Beijing 100190, China}
\affiliation{Collaborative Innovation Center of Quantum Matter, Beijing, China}
\author{Xianggang Qiu}
\affiliation{Beijing National Laboratory for Condensed Matter Physics, Institute of Physics, Chinese Academy of Sciences, Beijing 100190, China}
\affiliation{Collaborative Innovation Center of Quantum Matter, Beijing, China}
\author{Pengcheng Dai}
\affiliation{Department of Physics and Astronomy, Rice University, Houston, Texas 77005-1827, USA}
\author{Yi-feng Yang}
\affiliation{Beijing National Laboratory for Condensed Matter Physics, Institute of Physics, Chinese Academy of Sciences, Beijing 100190, China}
\affiliation{Collaborative Innovation Center of Quantum Matter, Beijing, China}
\author{Huiqian Luo}
\affiliation{Beijing National Laboratory for Condensed Matter Physics, Institute of Physics, Chinese Academy of Sciences, Beijing 100190, China}
\author{Shiliang Li}
\email{slli@iphy.ac.cn}
\affiliation{Beijing National Laboratory for Condensed Matter Physics, Institute of Physics, Chinese Academy of Sciences, Beijing 100190, China}
\affiliation{Collaborative Innovation Center of Quantum Matter, Beijing, China}

\begin{abstract}

We have systematically studied the nematic fluctuations in the electron-doped iron-based superconductor BaFe$_{2-x}$Ni$_x$As$_2$ by measuring the in-plane resistance change under uniaxial pressure. While the nematic quantum critical point can be identified through the measurements along the (110) direction as studied previously, quantum and thermal critical fluctuations cannot be distinguished due to similar Curie-Weiss-like behaviors. Here we find that a sizable pressure-dependent resistivity along the (100) direction is present in all doping levels, which is against the simple picture of an Ising-type nematic model. The signal along the (100) direction becomes maximum at optimal doping, suggesting that it is associated with nematic quantum critical fluctuations. Our results indicate that thermal fluctuations from striped antiferromagnetic order dominate the underdoped regime along the (110) direction. We argue that either there is a strong coupling between the quantum critical fluctuations and the fermions, or more exotically, a higher symmetry may be present around optimal doping.

\end{abstract}

% insert suggested PACS numbers in braces on next line

\pacs{74.70.Xa,72.15.-v}

%\maketitle must follow title, authors, abstract, \pac
\maketitle

The normal-state electronic states in many iron-based superconductors show strong in-plane anisotropic properties that break the fourfold rotational symmetry of the lattice due to the presence of nematic order \cite{FernandesRM14,ChuJH10,ChuJH12,YoshizawaM12,BohmerAE14,LuX14,GallaisY13,ThorsmolleVK16,HosoiS16}. The nematic order is typically accompanied by a structural transition at $T_s$ following an antiferromagnetic (AF) transition at $T_N \leq T_s$, except for FeSe where no AF order is found \cite{LesterC09,McQueenTM09}. Both the AF order and the structural transition disappear around the optimal doping level, indicating the presence of magnetic and/or nematic quantum critical points (QCPs). While there is increasing evidence that the magnetic QCP may not exist in many materials \cite{PrattDK11,LuoH12,LuX13,DingH15,TanG16}, the nematic QCP has attracted more and more interest since nematic quantum fluctuations may induce an attractive pairing interaction and thus enhance or even lead to superconductivity \cite{LedererS15,MetlitskiMA15,SchattnerY15}.

So far, the evidence for the nematic QCP is rather limited. It is shown that the nematic order may go through a zero-temperature order-to-disorder quantum phase transition as shown by elastoresistance \cite{ChuJH12,HosoiS16}, elastic constants \cite{YoshizawaM12,BohmerAE14}, and Raman scattering \cite{GallaisY13,ThorsmolleVK16}. The nematic susceptibility shows divergent behavior with the form of $(1/T)^\gamma$ around optimal doping \cite{ChuJH12,BohmerAE14,GallaisY13,ThorsmolleVK16,KuoHH16,HosoiS16}. However, the value of $\gamma$ is found to be 1, which suggests that the nematic QCP may result in a mean-field scaling. In the underdoped regime, the nematic susceptibility above the thermal transition $T_s$ should show Curie-Weiss-like behavior according to the Landau theory of second-order phase transitions. Therefore, quantum nematic fluctuations seem to show no characteristics distinguishable from thermal fluctuations, which is rather strange since one always expects that quantum and thermal fluctuations within the same system give rise to different critical properties. 

It has been well accepted that the nematic order in iron pnictides is an Ising-type order \cite{FernandesRM14}, which only gives the nematic signal along the in-plane (110) direction in the tetragonal notation. Since its thermal fluctuations may be treated as fluctuating nematic ``moments" pointing to both (110) and (1$\bar{1}$0) directions, applying pressure along the (100) direction results in zero signal due to the cancellation between the above two directions \cite{ChuJH12}. However, since nematic quantum critical fluctuations should manifest themselves in the charge channel due to their coupling to fermions, the nematic signal may be observed along the (100) direction in the resistivity measurement since the fermions may be scattered more isotropically around the QCP\cite{SchattnerY15}. Moreover, it is also possible that a higher symmetry may present around the QCP, as shown both theoretically and experimentally\cite{ZamolodchikovAB89,IsakovSV03,EsslerFHL04,FonsecaP06,ColdeaR10}. In any case, the signal observed along the (100) direction may be treated as a sign of nematic quantum fluctuations. 

In this Letter, we give a systematical study on the nematic fluctuations in the electron-doped BaFe$_{2-x}$Ni$_x$As$_2$ by measuring the resistance change under uniaxial pressure. Our measurements along the (110) direction confirm previous reports on the presence of a nematic QCP and the Curie-Weiss-like behavior in the underdoped regime \cite{ChuJH12}. On the other hand, a sizable signal of the pressure-dependent resistance along the (100) direction has been detected, which is prohibited in a simple Ising-type model. By comparing the results between two directions, we argue that most of the signal along the (110) direction observed in the underdoped samples comes from thermal fluctuations of the striped AF order. Therefore, nematic quantum critical fluctuations can be unambiguously measured along the (100) direction or the (110) direction in the overdoped regime.

\begin{figure}[tbp]
\includegraphics[scale=.5]{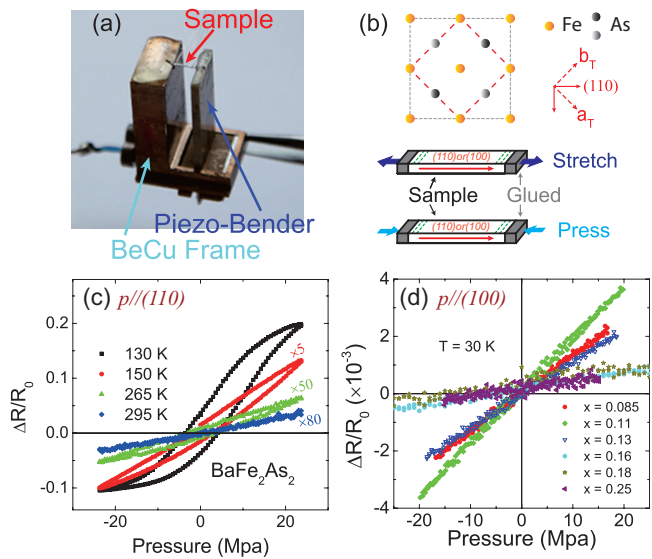}
\caption{(a)  The uniaxial pressure device, which is composed of a BeCu frame and a piezobender. The movement of the bender's top towards and away from the BeCu frame corresponds to pressing and stretching the sample, respectively. The force applied to the sample is proportional to the voltage applied to the piezobender. Since it is hard to measure the actual pressure, we define zero pressure as zero voltage of the piezobender. (b) Top view of the Fe-As block and sketch of the measurement. The tetragonal axes $a_T$ and $b_T$ are along the next-nearest Fe-Fe direction. Below the structural transition temperature $T_s$, stretching and pressing the sample along the tetragonal (110) direction above the saturation pressure will measure the resistivity along the orthorhombic $a_O$ and $b_O$ axes, respectively. (c) The pressure dependence of $\Delta R/R_0$ of BaFe$_2$As$_2$ along the (110) direction at several temperatures, where $R_0$ is the average resistivity at zero pressure and $\Delta R$ = $R(p)-R_0$, respectively. (d) The pressure dependence of $\Delta R/R_0$ along the (100) direction at 30 K for samples with different doping levels. 
}
\label{fig1}
\end{figure}

Single crystals of BaFe$_{2-x}$Ni$_x$As$_2$  were grown by the self-flux method that has been reported elsewhere \cite{ChenY11}. The orientations of the crystals were determined by x-ray Laue method and the samples were cut into thin plates along the desired directions by a diamond wire saw. The piezoelectric device to measure the resistivity under uniaxial pressure is shown in Fig. 1(a). The force applied by the piezobender upon voltage is calibrated by another uniaxial pressure device with spring \cite{LuX14}. The applied pressure $p$ is calculated by $F/S$, where $F$ and $S$ are the force and sectional area of the sample, respectively. Zero pressure should be achieved at a certain voltage since the piezobender can continuously change from pressing to stretching the sample [Fig. 1(b)], which indicates that the results shown in this paper are intrinsic in the limit of zero pressure. 

Figure 1(c) gives the results along the (110) direction for the $x=0$ sample at different temperatures. The positive slope suggests that the resistance change under pressure is due to the nematic order and its fluctuations \cite{ChuJH12} since the resistance in a normal metal should become smaller under pressure due to compression. At high temperatures, a fusiform hysteresis behavior is found due to the ferroelastic properties of the piezoelectric ceramic. The hysteresis loop becomes ferromagnetism-like below $T_s$, which has also been observed in the lightly underdoped samples. The asymmetric shape below $T_s$ may come from the residual strain that causes the zero pressure apart from zero voltage. In a ferromagnetic material, the hysteresis loop of the magnetization versus the magnetic field comes from the change of the magnetic domains. Similarly, the hysteresis loop below $T_s$ should be related to the AF or nematic domains. 

As discussed above, the same measurement along the (100) direction should give zero value in an Ising-type model. In fact, one expects a negative slope due to the compression of the sample. Figure 1(d) gives the pressure dependence of $\Delta R/R_0$ along the (100) direction at 30 K. Surprisingly, the sign of the slope is still positive, suggesting that the compression effect is negligible and the signal probably has the same origin as that along the (110) direction [Fig. 1(c)].

The response of the resistivity under the uniaxial pressure along the (110) direction in the absence of the nematic order gives a measure of the nematic fluctuations\cite{ChuJH12}. Above $T_s$, $R_a-R_b$ can expressed as $R(-P)-R(P)$ = 2 $\Delta R$ for $P > 0$ considering that the resistance depends linearly on pressure. One may use the slope of the resistivity change above $T_s$ to study the temperature dependence of the nematic fluctuations , i.e., $d(\Delta R/R_0)/dp \propto d\psi/dp$ where $\psi$ and $p$ are the resistivity anisotropy and pressure, respectively. While the fusiform hysteresis makes it hard to obtain the precise pressure dependence of the resistivity, its average slope is still a good approximation to $d\psi/dp$. Here, we define $\zeta$ as $d(\Delta R/R_0)/dp$ that is proportional to nematic susceptibility if the pressure is along the (110) direction. 

For the samples with $x \leq$ 0.12 [Fig. 2(a)], we can fit the high-temperature data of $\zeta_{(110)}$ with a simple Curie-Weiss-like function $A/(T-T')+y_0$, where $A$ and $y_0$ are temperature-independent constants, and $T'$ corresponds to the mean-field nematic transition temperature. These results are consistent with previous results obtained by measuring $d\psi/d\epsilon$ where $\epsilon$ is strain\cite{ChuJH12}, but no temperature-dependent parameter is introduced in our fitting, suggesting the reliability of the data. While it has been argued that a different temperature dependence may be observed for $d\psi/d\epsilon$ and $d\psi/dp$ \cite{ChuJH12}, our results suggest that the coefficient associated with the lattice in the Ginzburg-Landau free energy should have a very weak temperature dependence \cite{BohmerAE16}. $T'$ becomes zero around $x = 0.11$, where the nematic QCP should locate. In the heavily overdoped samples, the temperature dependence of $\zeta_{(110)}$ shows a broad hump peaked at $T_h$ as shown in Fig. 2(b). We note that $T_h$ may be related to the $T^*$ obtained in the elastic shear modulus measurement \cite{YoshizawaM12,BohmerAE14}, which suggests that the coupling between the nematic fluctuations and lattice indeed persists over the whole superconducting dome. 

The temperature dependence of $\zeta_{(100)}$ is shown in Fig. 2(c). While the signal in the underdoped samples shows an upturn close to $T_s$, they cannot be fitted by the Curie-Weiss-like function. The upturn may be associated with thermal fluctuations of the nematic order although we cannot totally rule out the possibility that it may come from the influence of $\zeta_{(110)}$ due to slight misalignment of the sample. The hump feature is also found in the overdoped samples, where $T_h$'s are similar to those determined from $\zeta_{(110)}$. Although $\zeta_{(100)}$ exhibit linear behavior in the $x=$ 0.11 sample, whether it may diverge at 0 K is unknown due to the presence of superconductivity. Comparing the doping evolution between $\zeta_{(100)}$ and $\zeta_{(110)}$ at low temperatures, a significant difference is that the former becomes maximum at $x =$ 0.11, while the latter continuously decreases with increasing Ni doping. 

\begin{figure}[tbp]
\includegraphics[scale=.43]{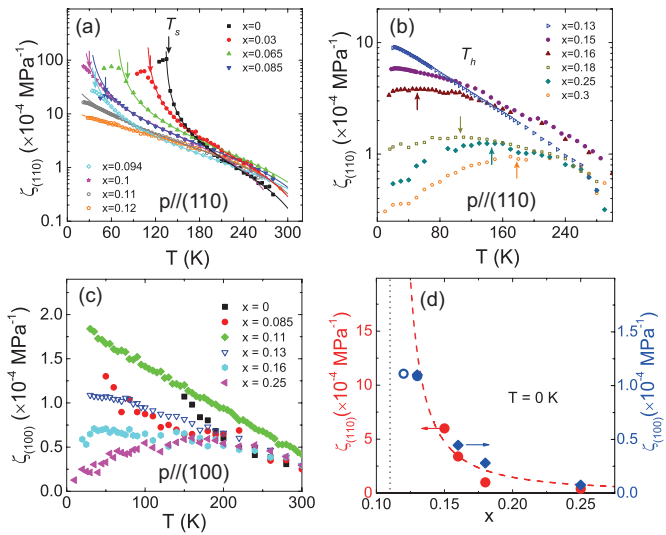}
\caption{ (a) Temperature dependence of $\zeta_{(110)}$ for $x \leq$ 0.12 in log scale. The solid lines show the Curie-Weiss-like fitting results as described in the text. A kinklike feature is observed in undoped and slightly underdoped samples, which seems to be associated with either $T_s$ or $T_N$ \cite{ChuJH12}. (b) Temperature dependence of $\zeta_{110}$ for $x >$ 0.12 where the hump position is labeled as $T_h$. (c) Temperature dependence of $\zeta_{(100)}$. (d) Doping dependence of $\zeta_{(110)}$ (red circles) and $\zeta_{(100)}$ (blue diamonds) at zero K in overdoped samples. Because of the presence of superconductivity, the values are extrapolated from the data by the eye. The dashed red line is the fitted result as described in the main text. 
}
\label{fig2}
\end{figure}

Figure 2(d) plots the doping dependence of $\zeta_{(110)}$ and $\zeta_{(100)}$ at 0 K for overdoped samples. While $\zeta_{(110)}$ is about ten times larger than $\zeta_{(100)}$, both of them show divergent behavior when $x$ approaches 0.11. We fit the doping dependence of $\zeta_{(110)}$ ($T = 0$ K) by $B(x-0.11)^P$, where $P$ is about -1.5 $\pm 0.2$. We note that while the error of $P$ may be larger due to the weak signal in the overdoped samples and the presence of superconductivity, it will not affect the fact that the value of $\zeta$ increases rapidly when approaching $x = 0.11$. These results suggest that both $\zeta_{(110)}$ and $\zeta_{(100)}$ in the overdoped samples are associated with the nematic disordered state at low temperatures. In other words, they have the same origin in the overdoped regime.

To rule out the possibility that the signal may come from the imperfect alignment of the sample, we give a simple estimation as follows. If the sample is misaligned $\theta$ degree  to the (100) direction, the force projected to (110) and (1$\bar{1}$0) directions will give a signal of $\zeta_{(110)}\sqrt{2}$sin$\theta$ along the (100) direction. The ratio of $\zeta_{(100)}/\zeta_{(110)}$ is less than 0.1 at low temperatures in the overdoped samples, which would suggest a $4^\circ$ of misalignment that is highly unlikely. Moreover, both the temperature and doping dependence of $\zeta_{(100)}$ in the underdoped regime suggest that the influence of misalignment should be negligible.

\begin{figure}[tbp]
\includegraphics[scale=.5]{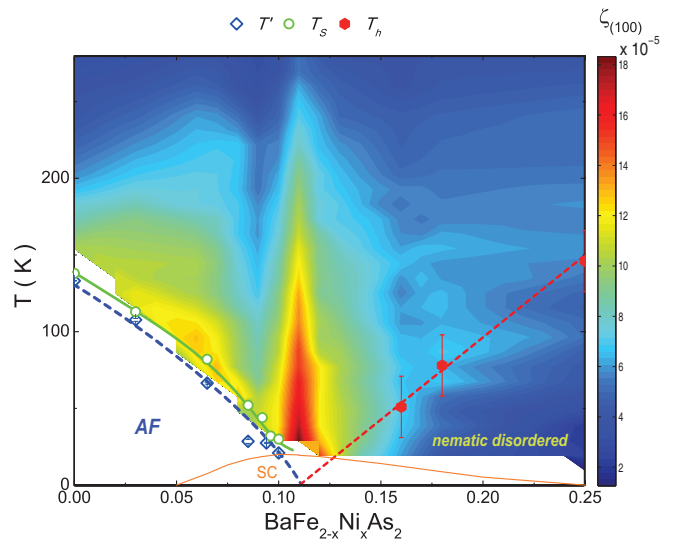}
\caption{ Nematic phase diagram of BaFe$_{2-x}$Ni$_x$As$_2$.  The green circles, blue diamonds, and red hexagons represent the actual structural transition temperature $T_s$ \cite{LuX13}, the fitted nematic transition temperature $T'$, and the hump temperature $T_h$, respectively. It should be noted that the values of $T_h$ determined from $\zeta_{(110)}$ and $\zeta_{(100)}$ are similar. The error bars of the $T’$ and $T_h$ are determined from the fitting error and artificially set to 20 K, respectively. The color map shows $\zeta_{(100)}$.
}
\label{fig3}
\end{figure}

Figure 3 gives the phase diagram of BaFe$_{2-x}$Ni$_x$As$_2$. The fitted mean-field nematic transition temperature $T'$ is always lower than the structural transition temperature $T_s$. The zero value of $T'$ around $x = 0.11$ suggests the disappearance of the nematic order. The zero value of $T_h$ extrapolated linearly from higher doping also happens at the same doping level, indicating that it is the crossover temperature that is associated with the nematic disordered phase. The decrease of both $\zeta_{(110)}$ and $\zeta_{(100)}$ with decreasing temperature distinguishes the nematic disordered phase from the quantum nematic fluctuations. The doping dependence of the positive $T'$ (more precisely, $T_s$ in the underdoped regime) and $T_h$ thus constitutes the funnel feature commonly found in a QCP system. 

The most prominent feature of the color map in Fig. 3 is that $\zeta_{(100)}$ becomes maximum right at the nematic QCP, which clearly suggests that it is directly associated with quantum critical fluctuations. In a recent elastoresistance study on the FeSe$_{1-x}$S$_x$ system \cite{HosoiS16}, the nematic signal along the (110) direction is found to be largest at the nematic QCP, which is different from the Co-doped Ba-122 system \cite{ChuJH12} and our results. Since an obvious difference between FeSe$_{1-x}$S$_x$ and Ba-122 systems is the lack of AF order in the former, the large thermal critical fluctuations present in the underdoped Ba-122 samples may be attributed to the fluctuations of striped AF order \cite{FernandesRM11,BreitkreizM14,GastiasoroMN14,WangY15}, which gives the same rotational symmetry breaking as the nematic order. This is consistent with the suggestion that no AF QCP presents in these materials \cite{PrattDK11,LuoH12,LuX13} and thus the AF order only contributes to thermal fluctuations along the (110) direction. 

The observation of nematic fluctuations along the (100) direction is unexpected since it is inconsistent with the Ising symmetry of the nematic order. While the main purpose of this Letter is to report the experimental results, we give a brief discussion on the origin of nonzero $\zeta_{(100)}$. The reason for the violation of Ising symmetry in our results may lie in the fact that the electrons in iron-based superconductors are itinerant. In an itinerant ferromagnetic system, both the quasiparticle properties and the dynamics of order parameters may be altered fundamentally around the QCP \cite{HertzJA76,MillisAJ93}. It has been argued that nematic order may fall into the same theory as a zero-momentum order \cite{LeeWC13}. A recent Monte Carlo simulation in a metal suggests that the quantum critical nematic fluctuations has a high degree of isotropy at the nematic QCP \cite{SchattnerY15}. Consequently, the pressure-dependent resistivity may also exhibit some degree of isotropy. In addition, we cannot rule out the possibility that a higher symmetry may be present around the QCP as suggested both theoretically \cite{ZamolodchikovAB89,IsakovSV03,FonsecaP06} and experimentally in the one-dimensional Ising chain system in the transverse field \cite{ColdeaR10}.  

In conclusion, our systematical study on the pressure-dependent resistance in BaFe$_{2-x}$Ni$_x$As$_2$ shows that while an Ising-type nematic order only allows nematic response along the (110) direction, a sizable signal is observed along the (100) direction. Both the doping and temperature dependence of $\zeta_{(100)}$ suggest that it is associated with the nematic QCP. Therefore, the measurement along the (100) direction provides an unambiguous detection on nematic quantum critical fluctuations. Our results ask further theoretical studies to fully comprehend the nematic QCP in iron-based superconductors.

This work is supported by the ``Strategic Priority Research Program（B）" of the Chinese Academy of Sciences (XDB07020300), the  Ministry of Science and Technology of China (No. 2012CB821400, No. 2015CB921302, No. 2016YFA0300502), the National Science Foundation of China (No. 11374011, No. 11374346, No. 11305257, No. 11322432, No. 11674406 and No. 11674372), and China Academy of Engineering Physics(No. 2015AB03). H. L. and Y. Y. are supported by the Youth Innovation Promotion Association of CAS. Work at Rice is supported by NSF DMR-1362219, DMR-1436006, and in part by the Robert A. Welch Foundation Grant No. C-1839(P. D.).

\bibliography{BFNA}

\end{document}